\documentclass[10pt,journal,compsoc]{IEEEtran}
\ifCLASSOPTIONcompsoc
  \usepackage[nocompress]{cite}
\else
  \usepackage{cite}
\fi
\usepackage{hyperref}
\hypersetup{
	colorlinks=true,
	linkcolor=black,
	filecolor=black,      
	urlcolor=blue,
	citecolor=black,
}
\ifCLASSINFOpdf
  \usepackage[pdftex]{graphicx}
\else
\fi
\usepackage{array}
\usepackage{multicol}
\usepackage{multirow}

\ifCLASSOPTIONcompsoc
 \usepackage[caption=false,font=footnotesize,labelfont=sf,textfont=sf]{subfig}
\else
 \usepackage[caption=false,font=footnotesize]{subfig}
\fi

\ifCLASSOPTIONcaptionsoff
 \usepackage[nomarkers]{endfloat}
\let\MYoriglatexcaption\caption
\renewcommand{\caption}[2][\relax]{\MYoriglatexcaption[#2]{#2}}
\fi
\usepackage{url}

\begin{document}
%
\title{A Survey on User-Space Storage and Its Implementations}
\author{Junzhe Li, Xiurui Pan, Shushu Yi, Jie Zhang
\IEEEcompsocitemizethanks{\IEEEcompsocthanksitem Junzhe Li is with the Department of
Electronics Engineering and Computer Science, Peking University. Email: allenli-@pku.edu.cn
\IEEEcompsocthanksitem Xiurui Pan is with the Department of Electronic Engineering, Tsinghua University. Email: panxiurui@outlook.com
\IEEEcompsocthanksitem Shushuyi and Jie Zhang are with the School of Computer Science, Peking University. Email:
firnyee@gmail.com, jiez@pku.edu.cn}}

\IEEEtitleabstractindextext{%
\begin{abstract}
The storage stack in the traditional operating system is primarily optimized towards improving the CPU utilization and hiding the long I/O latency imposed by the slow I/O devices such as hard disk drivers (HDDs).
However, the emerging storage media experience significant technique shifts in the past decade, which exhibit high bandwidth and low latency. These high-performance storage devices, unfortunately, suffer from the huge overheads imposed by the system software including the long storage stack and the frequent context switch between the user and kernel modes. Many researchers have investigated huge efforts in addressing this challenge by constructing a direct software path between a user process and the underlying storage devices. 
We revisit such novel designs in the prior work and present a survey in this paper. 
Specifically, we classify the former research into three categories according to their commonalities. We then present the designs of each category based on the timeline and analyze their uniqueness and contributions.
This paper also reviews the applications that exploit the characteristics of theses designs. Given that the user-space storage is a growing research field, we believe this paper can be an inspiration for future researchers, who are interested in the user-space storage system designs.  
\end{abstract}

\begin{IEEEkeywords}
Storage System, Solid-state disks, Non-volatile memory
\end{IEEEkeywords}}

\maketitle

\IEEEdisplaynontitleabstractindextext

%
\IEEEpeerreviewmaketitle

\IEEEraisesectionheading{\section{Introduction}\label{intro}}
The traditional storage media, such as hard disk drivers (HDDs), are commonly considered as slow I/O devices, whose performance is multiple orders of magnitude worse than the main memory \cite{mittal2015survey,zhang2015study}. 
To prevent the slow devices from stalling the execution of the user applications, a traditional operating system is usually split into user space and kernel space. While users execute their applications in the user space without caring about the explicit computer hardware, the kernel space is responsible for interacting with all the peripheral I/O devices. Although the user-kernel mode switch introduces additional execution time \cite{zhang2015nvmmu}, such overheads are fairly minor compared to the slow read/write latencies of the traditional HDDs. 

However, as the storage techniques shift, there emerge a couple of high-performance storage devices such as storage-class memory (SCM) \cite{SCM,SCMsurvey} and NVMe SSDs \cite{nvmewebsite}. Compared to HDDs, these new techniques significantly narrow down the performance gap between the storage and the memory. In addition, they exhibit brand new features that have never been unveiled by the stale storage devices. Specifically, SCM is usually comprised of non-volatile memory (NVM) \cite{SCM}. It achieves the read/write latency similar to the traditional DRAM. It also provides byte-granule data accesses to the users such that the user programs can directly access SCM via the standard load/store instructions. As a type of storage media, SCM also guarantees data persistency. 
On the other hand, solid state drives (SSDs) employ several dozens of flash dies, which can serve the I/O requests in parallel. It also employs a high-performance communication protocol, called non-volatile memory express (NVMe) \cite{nvmewebsite}, which is customized to exploit the internal parallelism of solid state drives. While NVMe SSDs are block devices, the accumulated throughput of the state-of-the-art NVMe SSDs are close to the commodity main memory.

While the I/O access latencies of the emerging storage devices decrease significantly, many prior work observe that the system software overheads have become the dominant performance bottleneck \cite{zhang2015nvmmu}.
For example, as representatives of SCM and NVMe SSDs, an Optane DC persistent memory module (PMM) and a ultra-low-latency (ULL) SSD decrease their read latencies to 100 ns \cite{izraelevitz2019basic} and 3 us \cite{cheong2018flash}, respectively. However, the software latency of user-kernel context switch is reported to be 0.5$\sim$2 us \cite{le2017latency}, 
which is close to or even longer than the I/O access latencies of SCM and NVMe SSD. In addition, the storage software stack within the kernel space usually performs multiple address translations and boundary checks, which consumes over 5 us software latency for each I/O request \cite{huang2015unified}. Therefore, the traditional computer system is not capable of unleashing the entire benefits of the emerging storage devices, due to the huge performance disparity between the existing system software and the emerging storage devices.


Considering that the involvement of system software in the I/O data path becomes the main reason for the performance degradation of the storage system, multiple prior work \cite{SPDKpaper,nvme,NVMeDirect,arrakis,2013arrakis,towardshighperformance,Aerie,zhu2018directfuse,yoshimura2019evfs,monetad,splitfs,zofs,devfs,crossfs} propose to provide user space with direct access to the underlying storage, referred to as \emph{user-space storage}. Specifically, many research \cite{SPDKpaper,NVMeDirect} concentrate on implementing user-space storage driver or I/O framework in order to grant the user applications full access to the underlying storage resources. Other prior studies \cite{arrakis,2013arrakis,towardshighperformance,monetad,simurgh,quill,vnvml2019,vnvml2020} consider to decouple the memory management and the system software such that the address translation and boundary check can be served directly from the user space. Furthermore, there are also multiple new designs of file systems \cite{zhu2018directfuse,ishiguro2012,yoshimura2019evfs,Aerie,Strata,splitfs,zofs,umfs,devfs,crossfs,fsp} to minimize the involvement of kernel in the file accesses. These work propose to construct user-space file systems, which can take over the tasks of the traditional file system.     
This paper gives a survey to summarize all the efforts in mitigating the penalty imposed by the traditional storage software stack including the innovative solutions for the storage driver, virtualization and file system.

\begin{figure}
    \centering
    \includegraphics[width=0.45\textwidth]{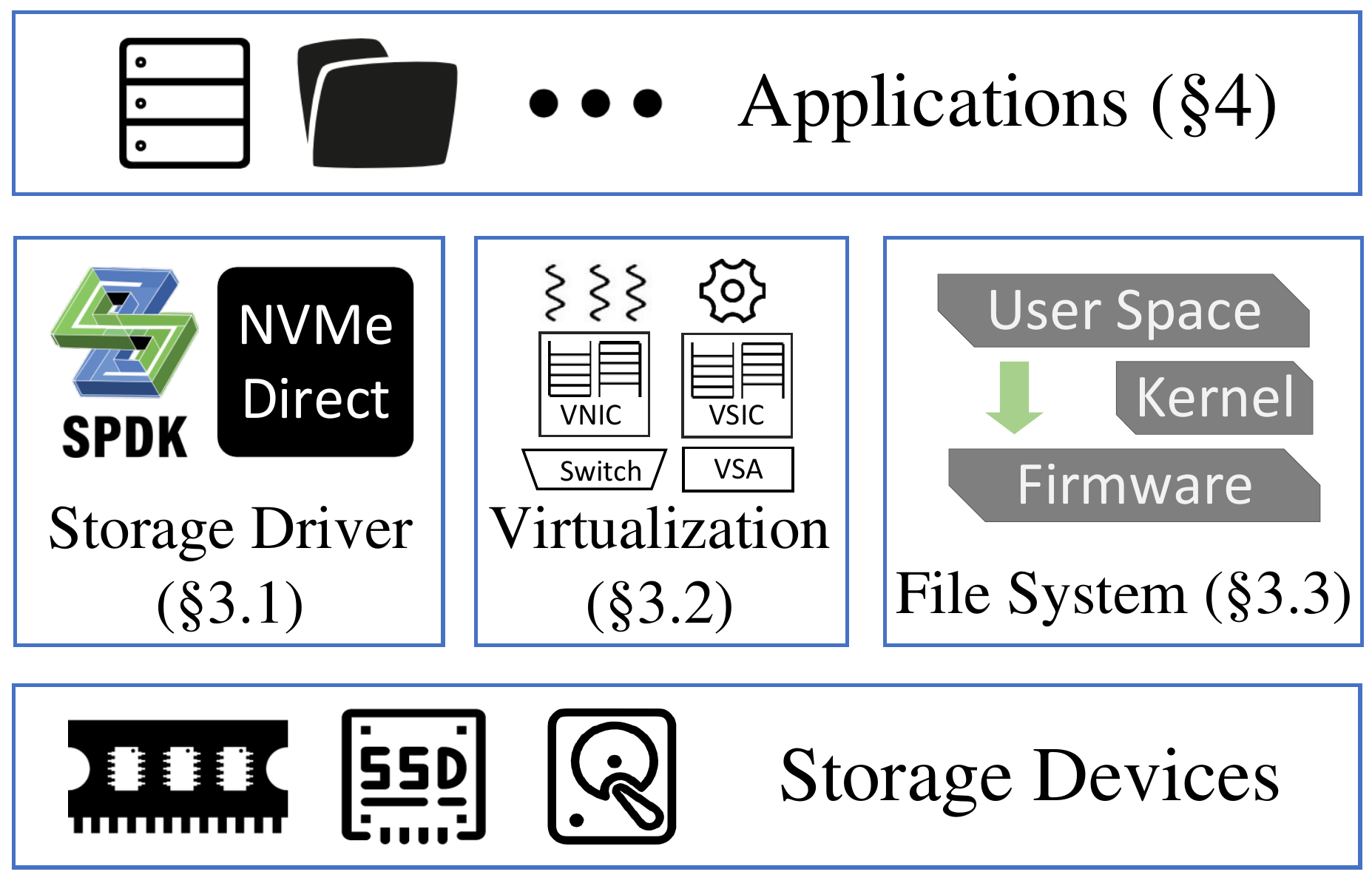}
    \caption{Three major system components of user-space storage designs and their applications.}
    \label{fig1}
\end{figure}

The remainder of the paper is organized as follows. Section \ref{bkground} describes the background and motivation for the hot research trend of user space storage at the moment. Section \ref{userspace} introduces the existing works that explore the designs of user space storage following the academic lineage and analyses the advantages and limitations of these works. Section \ref{application} presents the user-space applications. Lastly, Section \ref{conclusion} concludes this paper.

\section{Background}
\label{bkground}
\subsection{Emerging Storage Techniques}
\label{storagetech}
\begin{figure}
    \centering
    \includegraphics[width=0.48\textwidth]{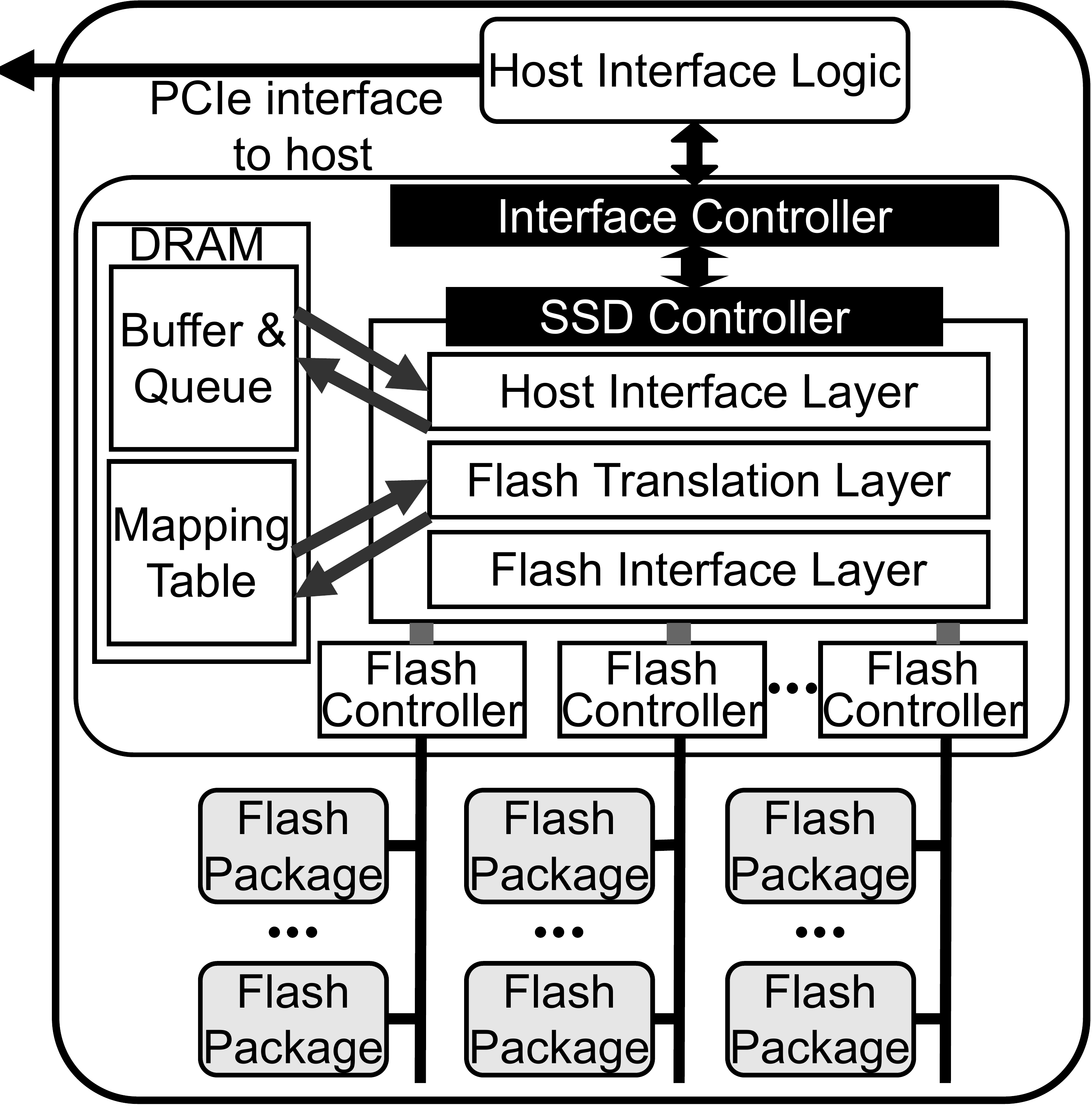}
    \caption{The structure of an SSD internal.}
    \label{fig:ssd}
\end{figure}

\noindent\textbf{NVMe SSDs.} 
As a replacement of the traditional hard disk drivers, the solid state drives (SSDs) have become the dominant storage media in the diverse application domains \cite{JIN20191,xu2015performance,xu2015performance2}. Compared to HDDs, SSDs can deliver significantly higher throughput by exposing their internal parallelism. Figure \ref{fig:ssd} shows the details of an SSD internal. SSDs typically consist of internal DRAM modules, a large number of flash packages, and several controllers and embedded cores over channel and system buses, which are connected to either MCH or ICH. Since the working frequency domains between host side hubs (MCH/ICH) and SSD device(s) are completely different, all I/O requests coming from the host-side hubs should be first buffered in SSD internal DRAM modules. The requests then again are transmitted to the data or cache registers of the underlying flash for back-end I/O services. To increase storage capacity and throughput, modern SSDs employ multiple channels, each containing a flash controller and a number of flash packages over its flash system bus such as ONFi \cite{grunzke2010onfi}, also referred to as ways. The low-level bandwidth of flash is around 70 MB/s \cite{jung2017simplessd}, which is far from capturing the bandwidth of the system bus or storage interface (4$\sim$8 GB/s \cite{gouk2018amber}). Thus, maximizing SSD internal parallelism is a key of designing modern high-performance SSDs. In practice, the SSD controller spreads the I/O requests across multiple channels and ways with four different levels of internal parallelism 
\cite{exploringsystemchallenge,jung2012evaluation}.
In addition, a customized communication protocol, referred to as Non-Volatile Memory express (NVMe), is used to enable users to take all the benefits of all levels of SSD internal parallelism 
\cite{zhang2020scalable}.
Specifically, NVMe allows the user applications to send I/O commands to upto 64K deep queues, each with up to 64K entries. Such massive deep queues allows the host to utilize as many hardware resources (e.g., threads) as possible for I/O accesses thereby maximizing the storage utilization. 

\noindent\textbf{Storage class memory.}
Storage class memory (SCM) has attracted a wide range of attention from both the academia and the industries as its non-volatile intrinsic, high density and low power consumption can benefit modern datacenters and high-performance computers. There are three standard incarnations of storage class memory including NVDIMM-N \cite{nvdimmn}, NVDIMM-F \cite{sainio2016nvdimm} and NVDIMM-P \cite{nvdimmp}. NVDIMM-N commonly consists of multiple volatile DRAM modules with a small piece of non-volatile memory (e.g., flash) for backup. On the other hand, NVDIMM-F directly integrates flash into a dual-inline memory module (DIMM). Similar to SSDs, NVDIMM-F provides a high memory capacity, but exposes a block interface to the users. As an ideal type of SCM, NVDIMM-P, such as Optane DC PMM \cite{optanpmm}, can offer byte-addressable persistency with DRAM-like performance. Thanks to these advantages, NVDIMM-P can be accessed via standard load/store instructions. In practice, enterprise servers (e.g., Intel Xeon scalable \cite{intelxeon}) employ NVDIMM-P with DirectAccess (DAX) \cite{DAX}, which brings the advantages of unprecedented levels of performance and data resiliency \cite{izraelevitz2019basic}.

\subsection{OS Storage Stack}
\label{userkernel}
\begin{figure}
    \centering
    \includegraphics[width=0.45\textwidth]{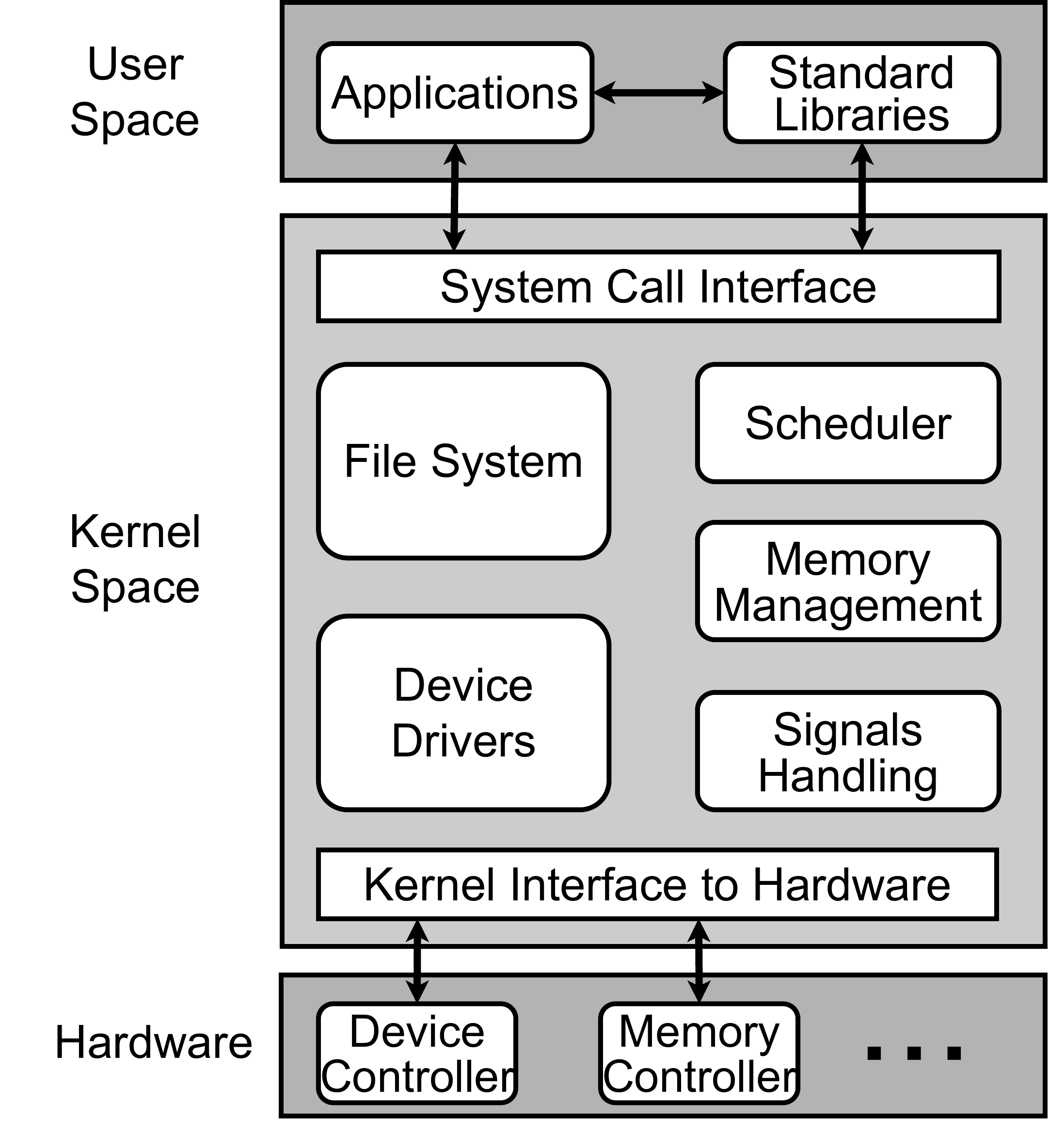}
    \caption{Software stack of UNIX-like operating system.}
    \label{fig:unix}
\end{figure}

Figure 
\ref{fig:unix}
shows the system structure of a representative operating system (UNIX \cite{unix}) from a user process to flash media. As shown in the figure, an I/O service is initialized by the user-space applications. It is then handled by the page cache, file system, multi-queue block layer and NVMe driver residing the kernel space.

\noindent\textbf{User-kernel interface.} 
User applications and runtime libraries typically reside in the user space, which can only access isolated resources. On the other hand, the kernel space is the core of the modern operating system and has permissions to access any hardware resources attached to the computer system. To get services beyond the permission range of user space, a user application can communicate with the OS kernel via a customized interface, referred to as \emph{system call} \cite{syscall}. The system calls specify the tasks that the user space can hand over to the OS kernel. The tasks include I/O request handling \cite{requesthandling}, CPU scheduling \cite{cpusche}, demand paging \cite{demandpaging} and page swapping \cite{pageswapping}. Once the OS kernel finishes the execution of a system call, it re-invokes the execution of user applications residing in the user space. This mechanism is referred to as user-kernel mode switch, which is usually accompanied by the context switch.

\noindent \textbf{Page cache.} Linux page cache stores page-sized chunks of files in the DRAM to speed up accesses to files on the underlying storage. The page-sized chunk (e.g., 4KB) is in practice managed by a radix tree, and if data is updated by a process, the target chunk is considered as a dirty page. When a user tries to read data over a file descriptor, the page cache receives the request at the beginning of the I/O process. It retrieves the file related information (\texttt{inode}) from the underlying file system and keeps it with the target page-sized chunk into DRAM.
Even though the page cache efficiently keeps the data by using a read-only page table entry or copy-on-write for sharing, the limited DRAM capacity cannot accommodate all pages requested by users. Thus, the page cache flushes dirty pages when the number of page dirty pages is greater than a threshold, referred to as the \texttt{dirty\_background\_ratio} or \texttt{dirty\_background\_bytes}. In addition, the page cache selectively writes the dirty pages whose period of life is longer than a timer that the user has configured. To guarantee data persistency, the user can make a system call that synchronizes all I/O operations or balances the dirty memory state. When this type of system call is used, the OS suspends the user process, generating file writes and performing the flush operations of the page cache. Since writing many dirty pages back to the SSD is a time consuming task, the threshold is periodically monitored by a kernel thread, \texttt{bdi\_fork\_thread}, and the thread creates a background task (\texttt{bdi\_writeback\_thread}) and periodically calls it to flush the dirty pages.

\noindent \textbf{File system.} Linux file system (e.g., EXT4) manages the storage space by abstracting all piece of information. The abstraction is guided by metadata (e.g., \texttt{inode}), which includes a file name, offset addresses indicating the beginning and end of a file, permissions, modifications and the last change. When the page cache requests I/O service due to flushing of a dirty page or a read of data from the system's backend, the file system retrieves the corresponding logical block address (LBA) and the length of the request in terms of the number of sectors (512 bytes). The file system then composes a block I/O structure instance, referred to as a \texttt{bio}, and calls a function (e.g., \texttt{ext4\_io\_submit}) to send it to the underlying block I/O layer.

\begin{table*}[!t]
\centering
\caption{Classification of relevant researches.}
\label{table:1}
 \begin{tabular}{ |p{2.7cm} p{2.7cm} p{2.7cm} p{2.7cm} p{2.7cm}|  }
 \hline
 Storage Driver & Virtualization  & FUSE & Comprehensive Design & Application \\
 Section \ref{driverlayer} & Section \ref{virtualization} & Section \ref{fuse} & Section \ref{comprehensivedesign} & Section \ref{application}\\
 \hline
 \hline
 SPDK \cite{SPDKpaper,spdkgithub,spdkwebsite}   & Arrakis \cite{arrakis,2013arrakis,towardshighperformance} & Ishiguro et al. \cite{ishiguro2012} &  Aerie \cite{Aerie} & RUMA \cite{ruma}\\
 NVMeDirect\cite{nvme}&  Moneta-D \cite{monetad}  & Direct-FUSE \cite{zhu2018directfuse}  & Strata \cite{Strata} & Breeze \cite{breeze}\\
  & Simurgh \cite{simurgh} &  XFUSE \cite{huai2021xfuse}&  SplitFS \cite{splitfs} & HyCache \cite{hycache}\\
    & Quill \cite{quill} & Son et al. \cite{designandevaluate2014,designandevaluate2015} &  ZoFS \cite{zofs} & Davram \cite{davram}\\
 & vNVML \cite{vnvml2019,vnvml2020}  & EvFS \cite{yoshimura2019evfs}& Kuco \cite{kuco} & DLFS \cite{deeplearning}\\
 &   & URFS \cite{tu2020urfs}   & UMFS \cite{umfs}& \\
&   &  & DevFS \cite{devfs}& \\
& & & CrossFS \cite{crossfs}& \\
& & & FSP \cite{fsp} & \\
\hline
 \end{tabular}

\end{table*}

\noindent \textbf{Block I/O layer.}  There are two types of block I/O layers. A conventional block I/O layer maintains a simple request queue as a request interface between the file system and the underlying driver/controller of the target interface. Since this introduces many performance issues due to a single lock on the queue management, a "multi-queue block layer", called \texttt{blk-mq}, is employed in most NVMe storage stack. Rather than a request interface, \texttt{blk-mq} actually composes an I/O request using a kernel data structure, called a \texttt{request}, by converting an incoming \texttt{bio} to a \texttt{request}. The \texttt{blk-mq}'s queues are created/released on a per-CPU or node basis. Thus, each CPU submits I/O requests into its own queue without contention on the single lock or interference with other CPUs. To further improve performance, \texttt{blk-mq} checks the target queue's entries and merges the incoming \texttt{request} into an existing entry, which is called aggregation. 

\noindent \textbf{Storage interface.} Under the block I/O layer, there is a storage interface driver, also known as the host block adapter. The implementation of this driver can vary based on the type of interface that the underlying SSD employs, but it is mostly related to compose commands being aware of the device-level registers or hardware map to communicate.  In cases of SATA/IDE, the target system employs a hardware controller (i.e., disk controller) to manage their storage interface protocol, so the interface driver usually handles I/O interrupt or system memory management. In contrast, in the case of NVMe, a kernel module (NVMe driver) directly accesses the PCIe bus over a memory mapped I/O and issues the request to the target SSD by composing an \texttt{nvme\_rw\_command}.


\subsection{Challenges} 
The designs of discrete kernel and user spaces can provide the traditional computing system with guaranteed isolation, security and consistency. However, this tight collaboration between the user and kernel spaces incurs frequent user-kernel mode switches, which introduces extra long latency. Specifically, \cite{huang2015unified} reports that the user-kernel mode switch costs 2$\sim$4 us latency. Considering that the I/O latency of the stale HDD-based storage system is millisecond-scale, the software latency caused by the context switch is relatively minor. However, the I/O latencies of the state-of-the-art storage devices (e.g., SCM and high-performance NVMe SSDs) have decreased to less than 10 us, which in turn exacerbates the software penalty. In addition, multiple prior research \cite{attackofthekiller,fasterthanflash,hwang2021rearchitecting,hildebrand2011revisiting} reveal that the storage stack, including virtual memory, file system, and storage drivers, imposes huge software overheads to the high-performance storage devices. For example, the storage stack requires multiple address translation and boundary checks, which cost around 10 us latency.


\subsection{Classification}
\label{classification}
To address the large overheads imposed by the tedious software stack of I/O services, there is continuous research on constructing a direct expressway between user space and the I/O devices. 
We review the prior works in the past decade aiming to address the software overhead issue in the storage system.
In general, most prior works concentrate on three system components: storage driver, virtualization technique, and file system, which are shown in Figure \ref{fig1}. Specifically, from the aspects of storage drivers, researchers propose to enable applications to directly access the underlying storage in user space by either moving drivers from the kernel to the user space or exploiting the unique features of the state-of-the-art storage drivers (e.g., NVMe drivers).
Virtualization-centric works map storage devices to user space and expose virtual interfaces to the applications, which are thereby permitted to access storage via such interfaces.
Lastly, the research of file systems propose to reassign tasks (e.g., metadata read/write and I/O permission checks) among the user space, the kernel space, and firmware so as to break free from the stale designs of traditional file system and put forward novel mechanisms to achieve user-space direct access.
Moreover, for various scenarios that call for fast access to storage (e.g., high performance computing), 
researchers have proposed several applications that are optimized for the emerging storage devices.
Table \ref{table:1} lists our classification of the relevant researches based on the core techniques or the application themes of each research.

\section{User Space Storage Designs}
\label{userspace}

In this section, we will review the three techniques (i.e., storage driver, virtualization and file system designs) and their related works in details. Section \ref{driverlayer} discusses works of storage drivers. 
Section \ref{virtualization} focuses on virtualization-related works. Section \ref{fuse} and \ref{comprehensivedesign} review file system designs, concentrating on both user-space-only file systems and comprehensive user-kernel-storage file systems.


\subsection{Storage Driver}
\label{driverlayer}

\noindent \textbf{NVMe engine in user space.}
One of the most representative NVMe engines in user space is \emph{Storage Performance Development Kits (SPDK)}, which is developed and released by Intel. SPDK includes a set of new tools and runtime libraries to eliminate the huge overheads imposed by the kernel I/O stack \cite{spdkwebsite,spdkgithub,SPDKpaper}.
Specifically, SPDK provides programmers with four software layers: application scheduling, storage services, storage protocols, and drivers. The first three layers allow users to design customized event schedulers, abstract storage resources, and develop diverse storage protocols, respectively.  
The last software layer (i.e., drivers) plays the fundamental role in mitigating I/O stack overheads in SPDK. SPDK places NVMe drivers in user space, which provides advantageous properties such as zero-copy and direct-access to NVMe SSDs for user-level applications. By exploiting the NVMe drivers, I/O requests from SPDK applications are processed in user space without going through the tedious I/O stack in kernel space.
However, implementing drivers in user space raises portability problems.
In particular, most applications relies on a uniform API (i.e., POSIX \cite{}) to access the entire storage stack. Unfortunately, SPDK is incompatible with POSIX APIs, making it difficult from being applied to a wide range of computer systems.
\noindent \textbf{NVMe engine in kernel space.}
The portability issue of SPDK is caused by the implementation of NVMe drivers in user space. To resolve this issue, Kim et al. \cite{NVMeDirect,nvme} proposes a new user-level I/O framework that mitigates the kernel I/O overheads while keeping the NVMe drivers in kernel space. 
The I/O framework is called NVMeDirect. It consists of three main components including an admin tool, an I/O queue management module, and a runtime library. The admin tool and I/O queue management module are implemented in kernel space while the runtime library is executed in user space.
Handling I/O requests in this framework requires collaboration of the three modules so as to achieve better performance in accessing NVMe SSDs. The user-space runtime library provides user-level APIs and invokes the kernel NVMe driver upon I/O requests from user space. The admin tools takes responsibility for controlling kernel drivers (i.e., NVMe drivers), which sets up the NVMe I/O queues, and managing the permission of I/O queues. The I/O queue management module handles such NVMe I/O queues and provides user-level applications with flexibility in choosing customized I/O policy (e.g., separation of reads and writes in a single thread).
NVMeDirect achieves higher I/O throughput and shorter read/write latency than SPDK and the kernel I/O in most cases \cite{nvme} while resolving the portability issue in SPDK. 

\subsection{Virtualization}
\label{virtualization}
\begin{figure*}
    \centering
    \subfloat[Storage Virtualization (e.g., Moneta-D \cite{monetad}).]{\includegraphics[width=0.32\textwidth]{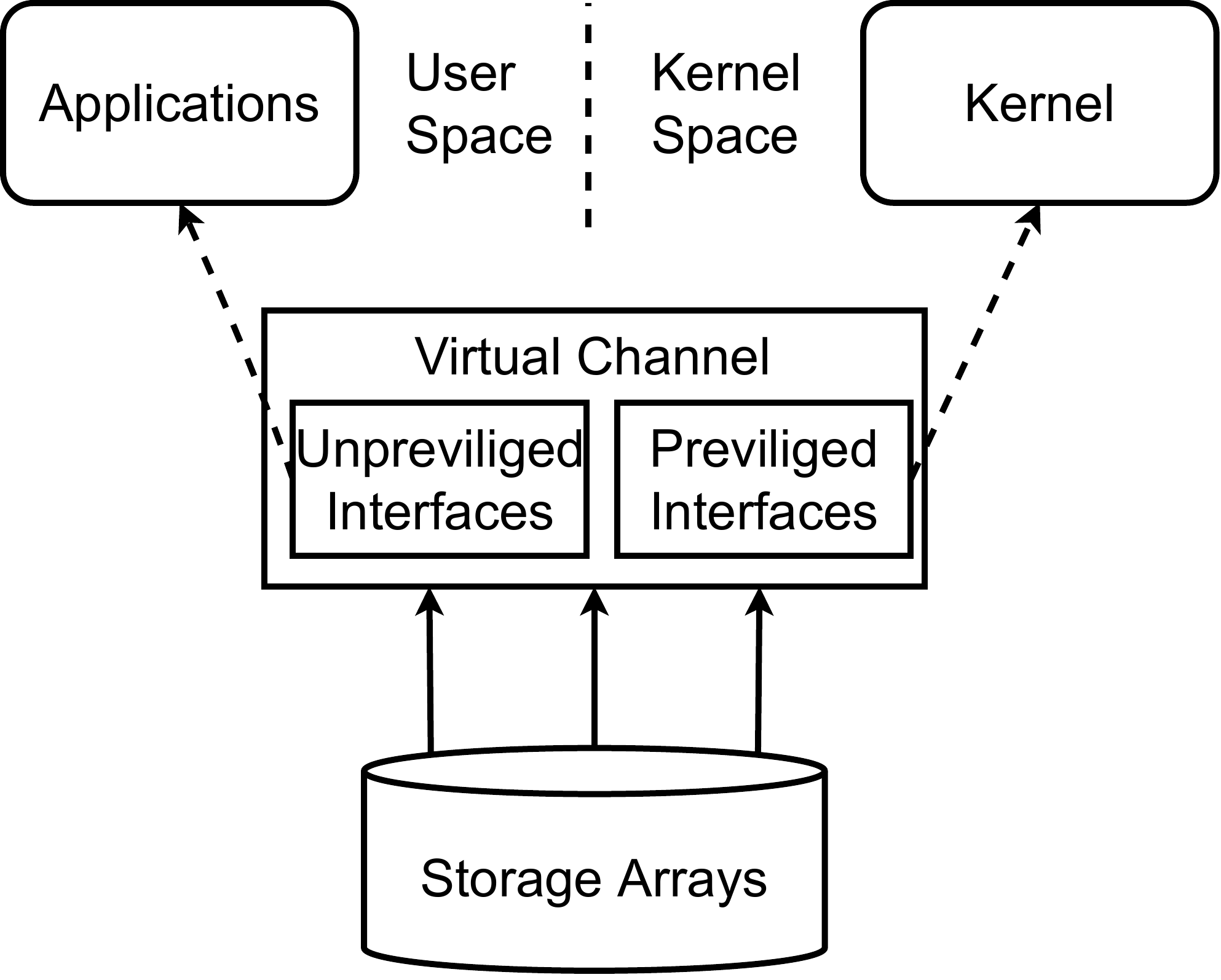}}
    \hfill
    \subfloat[Device Virtualization (e.g., Arrakis \cite{arrakis}).]{\includegraphics[width=0.32\textwidth]{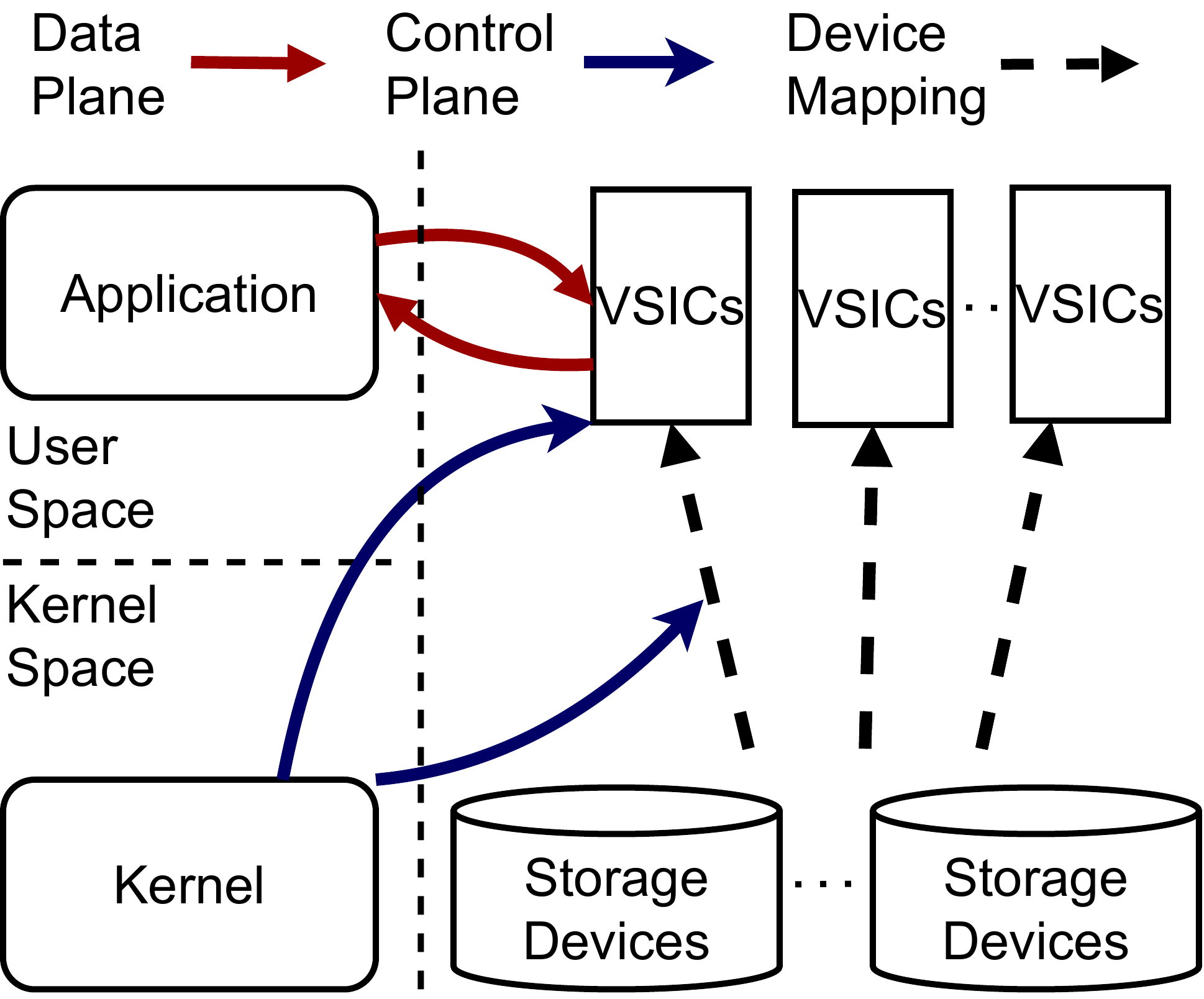}}
    \hfill
    \subfloat[NVM virtualization with the user-level library (e.g., Quill \cite{quill}, vNVML\cite{vnvml2019}).]{\includegraphics[width=0.32\textwidth]{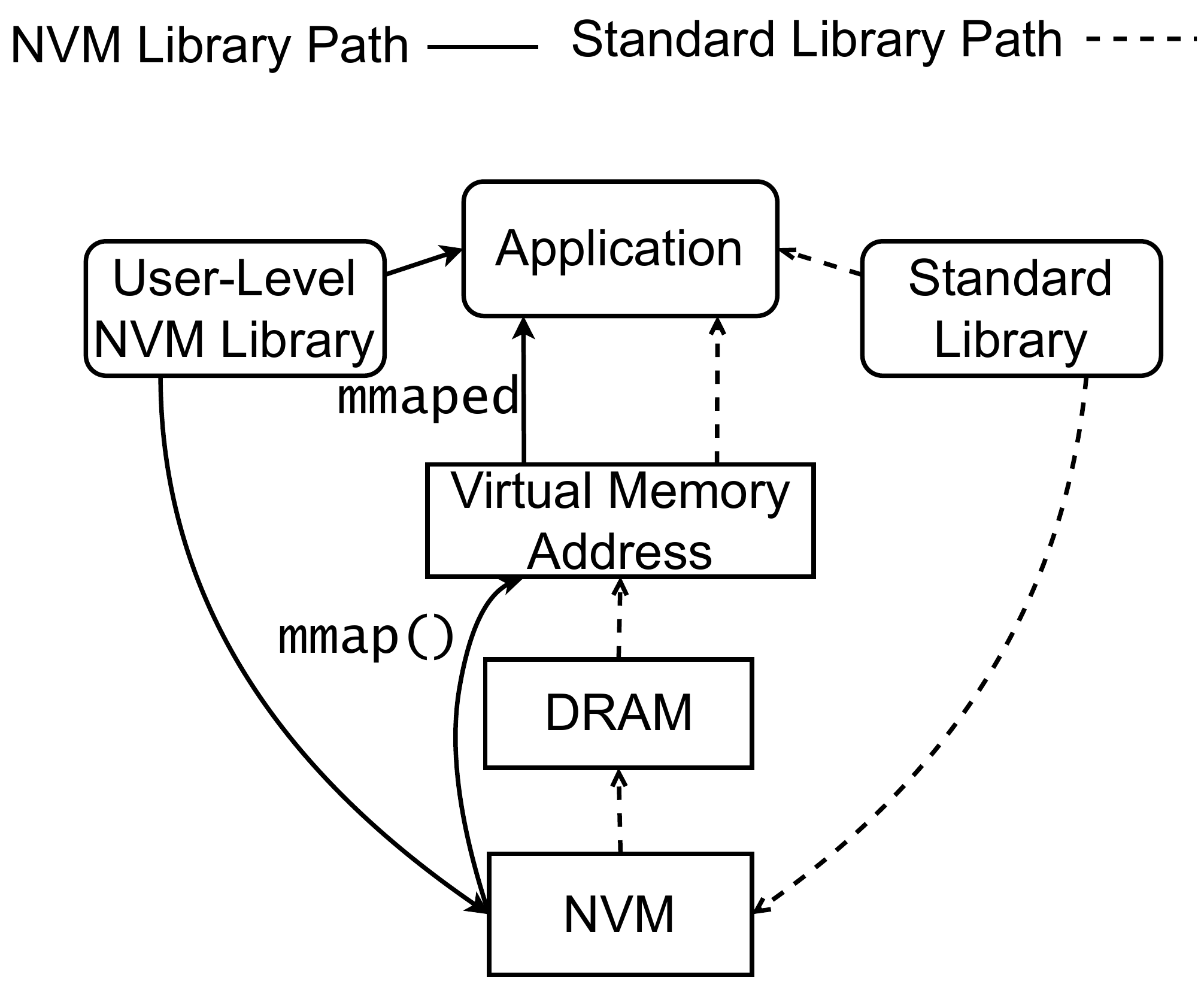}}
    \caption{Three major types of virtualization.}
    \label{fig:virtualization}
\end{figure*}

The traditional storage system suffers from software overheads such as context switch and metadata management. In order to reduce such overheads and also to limit kernel involvement, researchers have paid great attention to the virtualization technique \cite{artiaga2010using}. Based on the specific storage component that is being virtualized in each work, we divide prior works on virtualization into three major groups: device virtualization, storage virtualization and NVM virtualization. We show the general system structures of each virtualization technique in Figure \ref{fig:virtualization}.

\noindent \textbf{Virtual Interfaces}
\noindent \textbf{Storage virtualization.}
Caulfield et al. \cite{monetad} proposes Moneta-Direct (Moneta-D), a new type of storage architecture that integrates storage virtualization into its core design. 
It employs kernel only to process necessary management operations as the control plane and provides user space with direct and concurrent accesses to the storage devices. 
The key design to achieve such goals is the virtual channels, which are virtual interfaces provided to user space and are directly mapped to storage pages.
A virtual channel consists of both privileged and unprivileged interfaces. The former are exposed to the kernel for management and permission checks of each channel while the latter are exposed to user space, enabling it to directly access storage devices via the mapping to the storage pages. Moreover, multiple channels can serve I/O requests simultaneously thereby realizing concurrent access to the underlying storage.

Peter et al. \cite{arrakis,2013arrakis,towardshighperformance} proposes a specific device model named Arrakis for virtualized I/O, which abstracts the underlying hardware devices as virtual device instances in user space.
Specifically, Arrakis presents the storage devices as virtual storage interface cards (VSICs), which function in the same way as their corresponding physical devices from the perspective of users. 
Arrakis also proposes to handle user-level I/O requests efficiently by splitting the roles of operating systems into data plane and control plane. The operations of data plane are associated with data accessing, such as asynchronous reads and writes. The control plane, on the other hand, manages the core tasks of a system for reliability and security, such as I/O permission checking and hardware resource allocation.
For control plane, the device drivers in the kernel manage VSICs and the physical devices via a set of operations such as device creation and destruction. These operations do not interfere with data-plane operations during runtime.
For data-plane operations, user-level applications can directly issue I/O requests to VSICs during the execution, which, with the support of hardwares (e.g., DMA controller), directly associates with underlying storage.
To this end, thanks to the split of data plane and control plane, user-space application attains direct access to the underlying storage devices without the involvement of the kernel. 

\noindent\textbf{NVM virtualization.} 
\emph{device virtualization} and \emph{storage virtualization} exploit the virtualization of storage devices for user-space direct access without specifying a particular kind of storage media. In contrast, other researches focus solely on the virtualization of non-volatile memory (NVM) from the aspects of file systems and user-level libraries.

Moti et al. \cite{simurgh} designs a user-space file system named \emph{Simurgh}, which bases its core design on virtualizing NVM.
Considering that NVM achieves similar performance to DRAM and is byte-addressable,
Simurgh directly maps NVM into the address space of each application without employing DRAM to cache data and metadata from NVM. In this way, neither data copy between NVM and DRAM nor data buffering in DRAM is necessary. Thus, NVM can be accessed directly in user space without DRAM as the medium. 
Since the metadata is cached in DRAM, it is possible for metadata to be accessed concurrently by independent processes.
Moreover, Simurgh adds two additional instructions (i.e., protected jump and return) to the CPU ISA to securely execute user-space functions in privileged mode, which in turn reduces the kernel's involvement during runtime. 

Aside from embedding NVM mapping in file system design (e.g., Simurgh), researchers have also proposed more light-weighted tools (i.e., user-level libraries) for NVM mapping. 
Eisner et al. \cite{quill} proposes Quill, a user-level library specified for accessing NVM block devices. Quill targets at addressing the challenges imposed by the stale memory management mechanism (i.e., paging). 
Paging is a memory management technique, which retrieves data from secondary storage devices (e.g., SSD, NVM) and stores them to the main memory (e.g., DRAM) for further usage. 
However, considering that the performance behaviors of NVM (i.e., latency and bandwidth) are comparable to DRAM, data copy from NVM to DRAM cannot benefit from the high speed of DRAM, thus making paging inefficient for NVM.
Observing this feature, Quill tries to avoid paging of NVM block devices. To this end, since NVMs can be cached inside CPUs, they are organized as pages in CPUs' physical address space directly. When a user-level application accesses a file, Quill takes it over and uses the \texttt{mmap()} function to map the physical pages of NVM devices to the virtual address space of user-space applications. The files stored in NVM are permitted to be accessed directly by the application, thus eliminating the necessity of paging.

Rather than solely focusing on the latency and bandwidth properties of NVM, Chou et al. \cite{vnvml2019,vnvml2020} designs a user-level library named vNVML, which explores utilization of other features of NVM.
vNVML exploits the byte-addressable feature of NVM by providing a conventional file interface \emph{mmaping}, which enables applications to access data through byte-level load/store instructions.
Moreover, vNVML extends the available physical storage of virtual NVM by integrating DRAM and underlying storage devices (e.g., SSDs) into its design. 
vNVML leverages DRAM and NVM as the cache to the underlying storage devices. 
Data read from the underlying storage are cached in DRAM, and NVM functions as both the log buffer and the write cache.
When the user-level application issues read or write requests to the virtual NVM, it is actually accessing the cache to underlying storage, which owns much larger physical storage space than NVM. 
Therefore, the virtual NVM is much larger from the users' perspective.

\subsection{User-Space File Systems}
\label{fuse}
While the prior work of the storage driver and the virtualization techniques enable the user-space application directly access data from storage, the kernel still plays a significant role in the entire computing system. Specifically, the kernel needs to control the behaviors of storage drivers and maintain the mapping from storage to user-space address spaces.
Since file systems play an essential role in accessing underlying storage, other than simply concentrating on one particular aspect of the storage system, such as the storage driver or the virtualization techniques, researchers have committed to redesigning file systems so as to minimize the kernel's involvement in data accessing.

General file systems reside in kernel space. User applications can interact with the file systems via customized system calls (i.e., open). 
However, developing file systems in kernel can be complicated and challenging because kernel codes are deeply coupled and its architecture is complex.
Therefore, user-space file systems have attracted researchers attention since they are easier to develop.
In this subsection, we mainly review the works focusing on user-space file systems to obtain better performance in accessing storage devices, including the widely used framework called Filesystem in Userspace (FUSE), FUSE-based designs and other user-space file systems, which focus to one particular types of storage media (e.g., NVM and NVMe SSDs).

\noindent\textbf{FUSE overview.}
To explore new paths for storage accesses, Filesystem in Userspace (FUSE) provides users with fundamental software interfaces to develop customized file systems in user space \cite{Kantee2007ReFUSEU,Narayan2010UserSS,toolkit}, minimizing interference from the kernel.

Specifically, FUSE consists of modules residing in both the user and kernel spaces. The kernel module is registered as a file system driver also named \emph{fuse}. It manages kernel operations with the virtual file system (VFS), an interface that supports concrete file systems.
The user-space modules, including the \emph{libfuse} library and the \emph{fuse} daemon, are responsible for setting up the file system in user space and handling file system calls. 
Using FUSE to implement a new file system requires users to write a handler program, which specifies the behavior of the file system in responding to read/write requests from user space and links such program to the \emph{libfuse} library. 
After FUSE is mounted, this handler program is registered by the kernel module (i.e., \emph{fuse}) for handling runtime data requests.

\begin{figure}
    \centering
    \includegraphics[width=0.47\textwidth]{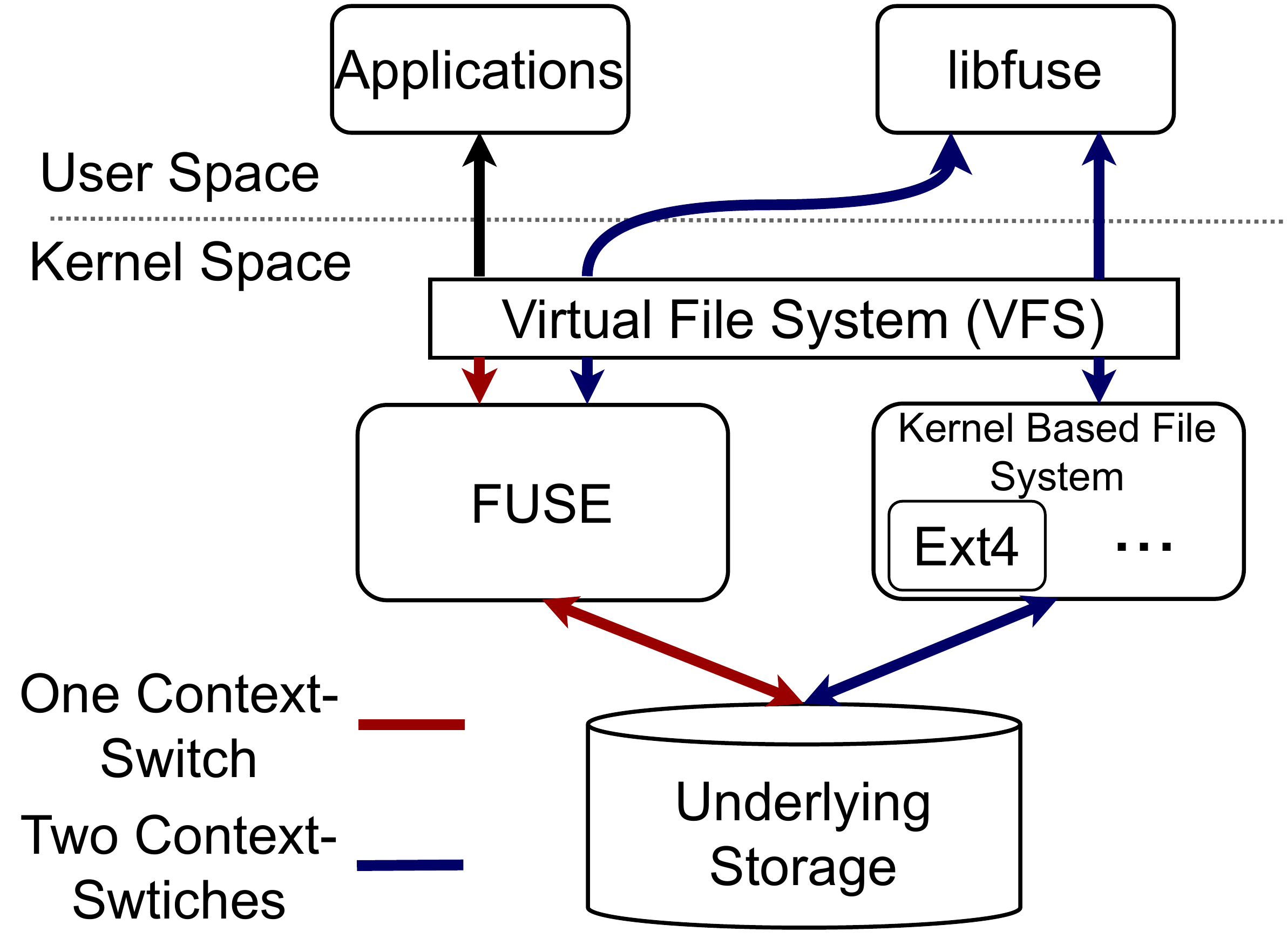}
    \caption{Execution of FUSE system calls.}
    \label{fig:fuse}
\end{figure}

The process of executing a system call in FUSE can be regarded as a client-server model, where the kernel \emph{fuse} module is the client and the \emph{fuse} daemon in user space is the server.
When a system call, such as \texttt{read()} or \texttt{write()}, is issued from user space,
the VFS first invokes the default handler in the \emph{fuse} to fetch data in the page cache.
The requested data is returned directly to user space if found in page cache.
Only one user-kernel context switch is required here. If not found, the system call is thereafter re-directed to the user-space module \emph{libfuse}. 
\emph{libfuse} invokes the user-defined handler in the kernel, which fetches the data according to the specification of the user and returns the fetched data back to user space. 
In this process, two user-kernel context switches are needed, as shown in Figure \ref{fig:fuse}. 
The kernel forwards the system call to \emph{libfuse}, which later invokes the handler in the kernel, and the kernel returns the requested data back to the user space. 

By providing the software interface (i.e., \emph{libfuse}), FUSE allows users to develop customized file systems conveniently. 
Moreover, FUSE moves parts of code execution from kernel to user space. 
Such reduction of code execution in kernel would reduce the possibility of kernel crashing. 
In addition, with \emph{libfuse} residing in user space and easy to be deployed in different environments, it is effortless to transplant FUSE-based file systems from one environment to another, which in turn makes FUSE achieve compatibility.

\noindent\textbf{FUSE modification for better performance.}
While FUSE enjoys many benefits as mentioned above (e.g., compatibility), kernel is massively involved in FUSE framework (e.g., VFS in handling system calls), which generates side effects. 
Specifically, when the user-defined handler is invoked, FUSE requires extra user-kernel context switches to transfer control between the handler and the \emph{fuse} daemon, which introduces non-negligible software overheads. 
In addition, extra memory copies, frequent page cache misses, and potential dead locks all lead to the performance degradation of FUSE \cite{rajgarhia2010performance}. 
Therefore, although FUSE-based file systems perform well in some scenarios (e.g., web-server workload), it is not suitable for metadata-intensive workloads (e.g., creating or deleting files over many directories) \cite{vangoor2017fuseornot}.
To minimize the overheads introduced by FUSE framework, several modifications to FUSE have been proposed.

Ishiguro et al. in 2012 \cite{ishiguro2012} modifies the kernel module of FUSE to achieve performance improvement. They redesign a mechanism that ports the \emph{fuse} daemon to the kernel and modify the kernel \emph{fuse} module to adapt to this mechanism. 
Thus, all the works that used to be done in the user space are handled by the kernel. This mechanism avoids redundant context switches and memory copies. 
Zhu et al. \cite{zhu2018directfuse}, on the other hand, focused on different types of FUSE. They describes a FUSE-based framework called Direct-FUSE to support multiple backend file systems without trapping into kernel. 
Direct-FUSE divides the FUSE framework into three layers including bottom, middle and top layers. 
The bottom layer (also known as the backend services) provides operations of multiple file systems while the middle layer provides unified file system interfaces for various backend services. 
Lastly, the top layer receives file system operations and identifies the corresponding backend service for each operation. 
All the layers of Direct-FUSE are in user space, thereby eliminating non-trivial overheads of user-kernel context switches.
Apart from minimizing the heavy overheads imposed by context switches, 
Huai et al. \cite{huai2021xfuse} puts forward a new user-space file system infrastructure called XFUSE. XFUSE targets at improving FUSE performance by adapting FUSE to modern multi-core CPU systems and the emerging storage devices (e.g., NVM). To improve the parallelism, XFUSE enables multiple file system daemon threads handling different requests in parallel, thereby increasing the scalability in multi-core systems.
To meet the speed of fast storage devices, it dynamically adjusts the period of busy-waiting according to the I/O requests, thus avoiding the unnecessary waits and improving the I/O throughput.
\noindent \textbf{User-space file systems on specific storage media.}
Though FUSE framework has gained great popularity in user-space file system development, it can not satisfy all needs raised by the emerging storage devices (i.e., NVM). Since NVM calls for faster and lightweight I/O stack, the prior work have employed new frameworks in developing user-space file systems to meet such demands.
To be specific, Son et al. \cite{designandevaluate2014,designandevaluate2015} observes the performance disparity between the slow I/O stack and the fast NVM. They propose a new user-level file system that aims to bridge this performance gap.
They provide users with a byte-grained interface to submit I/O requests and construct a user-space file system to serve storage I/O accesses.
The proposed scheme bypasses the bulky traditional I/O layers such as VFS, generic block layer, and page cache, which enable user applications to read/write byte-grained data from/to NVM directly thereby reducing I/O latency.

NVMe protocol has become the dominant storage interface, which can deliver low-latency and highly parallel access to the underlying NVM media (i.e., SSDs). NVMe-based SSDs bring about many advantages, including massive I/O queues and faster speed than traditional SSDs. These advantages have raised new challenges to the design of user-space file systems since they play the major role in accessing NVMe SSDs.
Yoshimura et al. \cite{yoshimura2019evfs} and Tu et al. \cite{tu2020urfs} propose EvFS and URFS, respectively. They both are user-level file systems that aim to fully exploit the advantages of NVMe SSDs.
EvFS employs SPDK \cite{SPDKpaper} to support various NVMe devices.
Specifically, EvFS adopts the event-driven execution model in SPDK for file I/O processing, targeting at the communication between users and the page cache in EvFS. 
Users submit the I/O operations as events to invoke the user-level NVMe driver in SPDK directly, which highly utilizes the bandwidth of NVM with few user threads and provides low I/O latency.
URFS, on the other hand, explores the parallelism provided by NVMe. It utilizes a series of memory sharing and protection mechanisms to share storage accesses among multiple I/O processes.
Applications that are executed upon URFS can share multiple NVMe SSDs via a library named \emph{fslib}, which provides POSIX-like file system APIs for the convenience of developers. 

\subsection{Comprehensive File System Design}
\label{comprehensivedesign}
\begin{figure*}
    \centering
    \subfloat[Client-Server model (e.g., Aerie \cite{Aerie}).]{\includegraphics[width=0.45\textwidth]{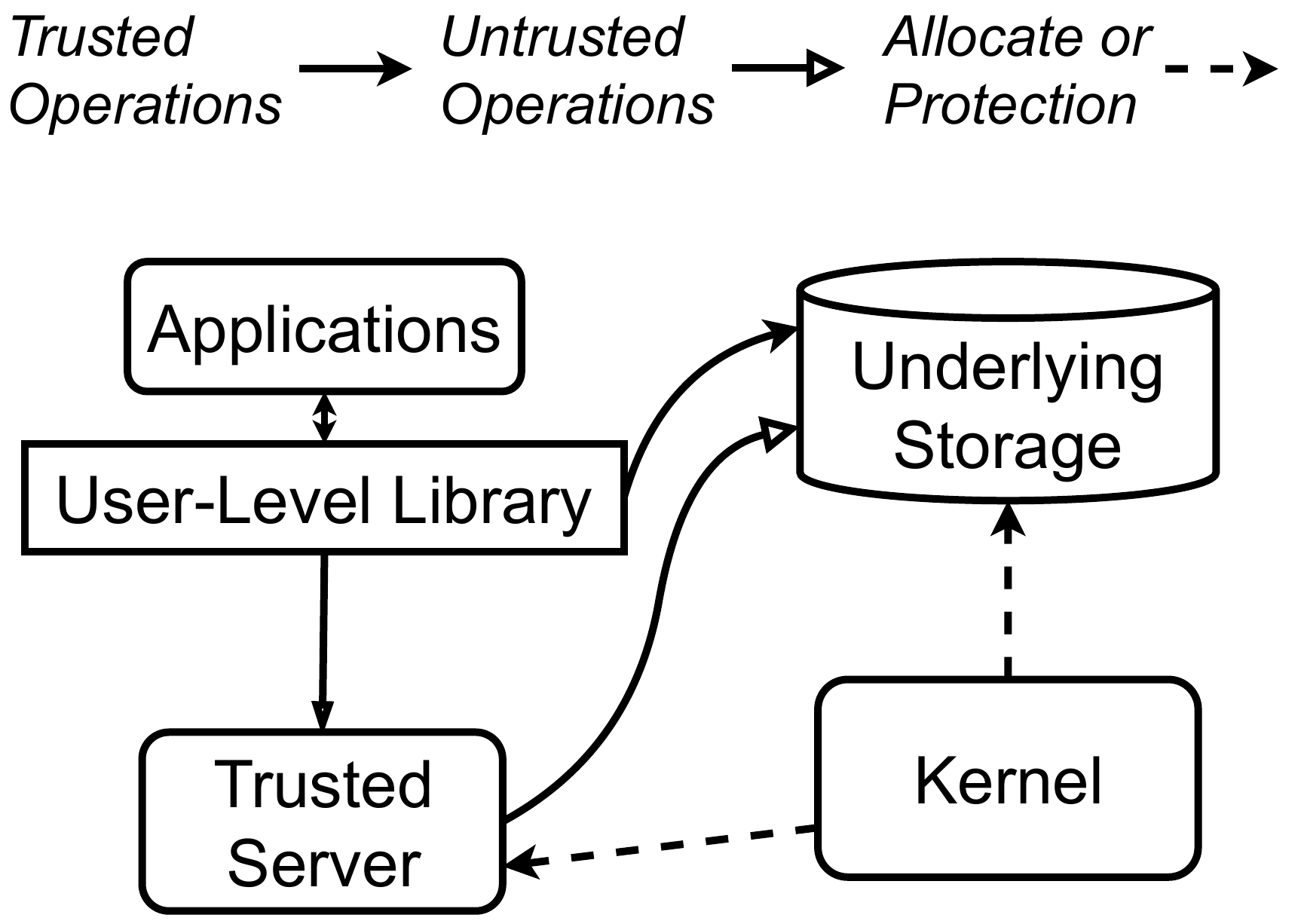}} \hfill
    \subfloat[User-Kernel split model (e.g., Strata \cite{Strata}, SplitFS \cite{splitfs}, ZoFS \cite{zofs}).]{\includegraphics[width=0.45\textwidth]{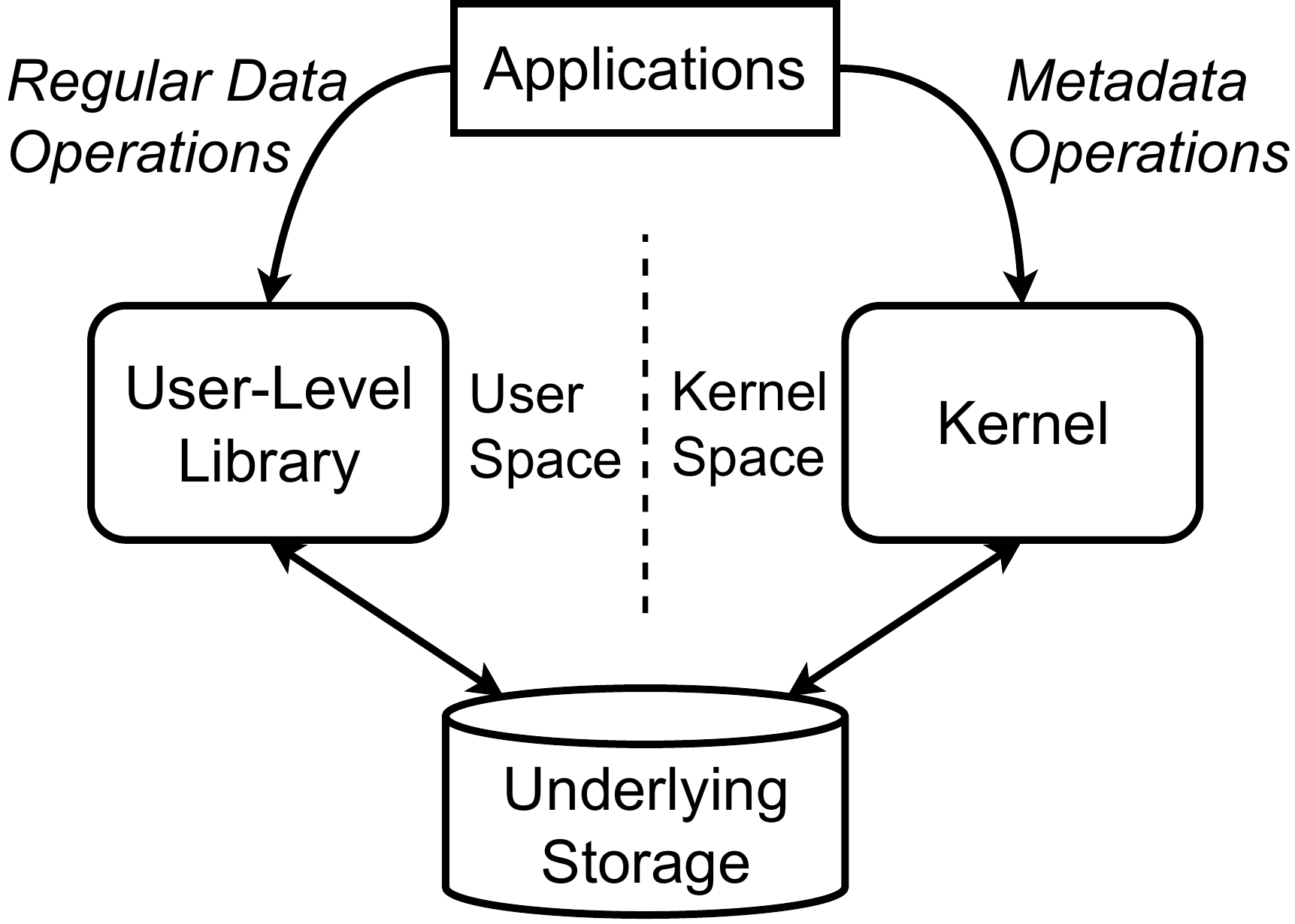}}
    \caption{Two representative models of kernel being the control plane.}
    \label{fig:control}
\end{figure*}

\begin{figure*}
    \centering
    \subfloat[FirmFS (e.g., in DevFS \cite{devfs}, CrossFS \cite{crossfs}).]{\includegraphics[width=0.47\textwidth]{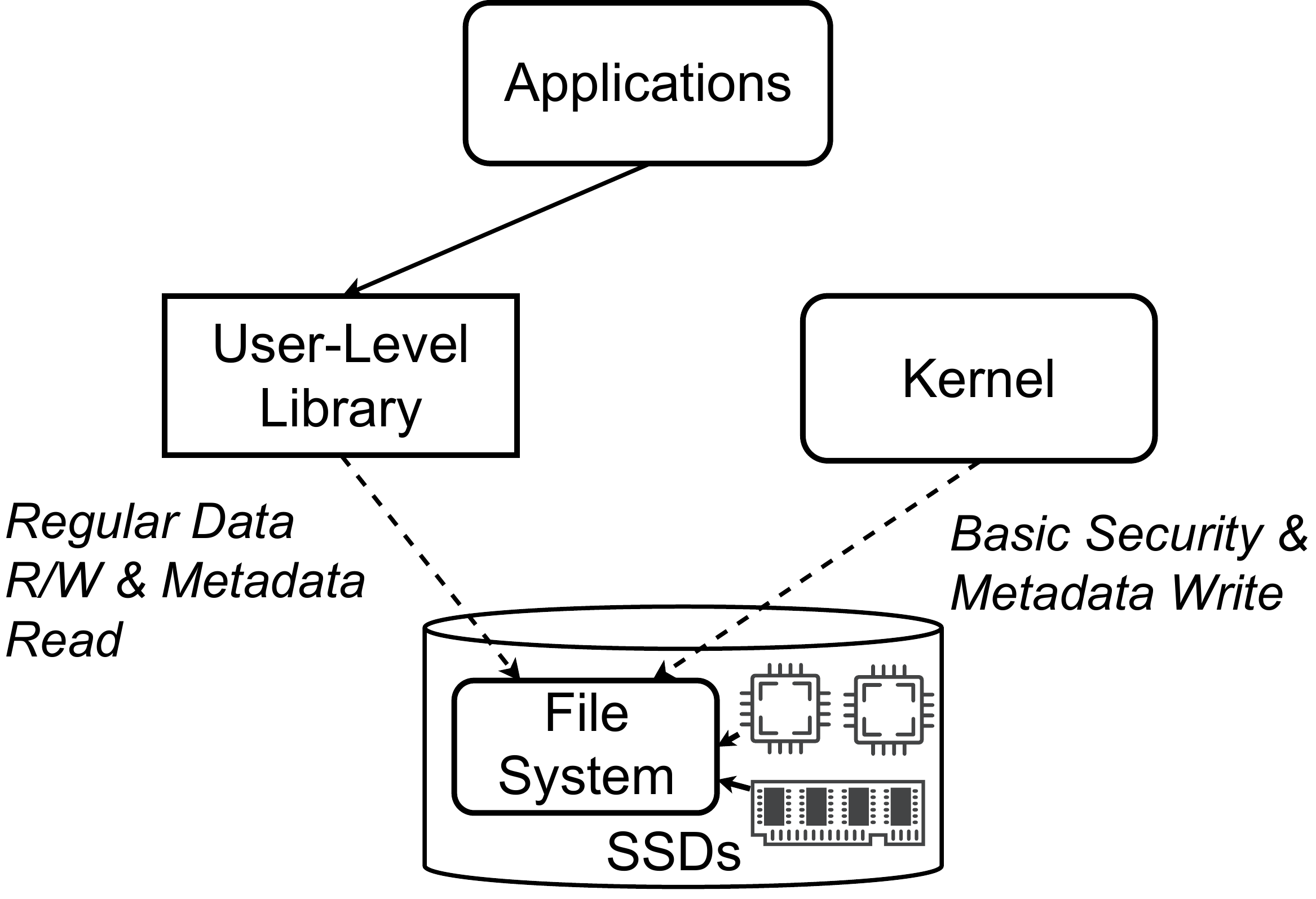}}\hfill
    \subfloat[File system as processes \cite{fsp}.]{\includegraphics[width=0.47\textwidth]{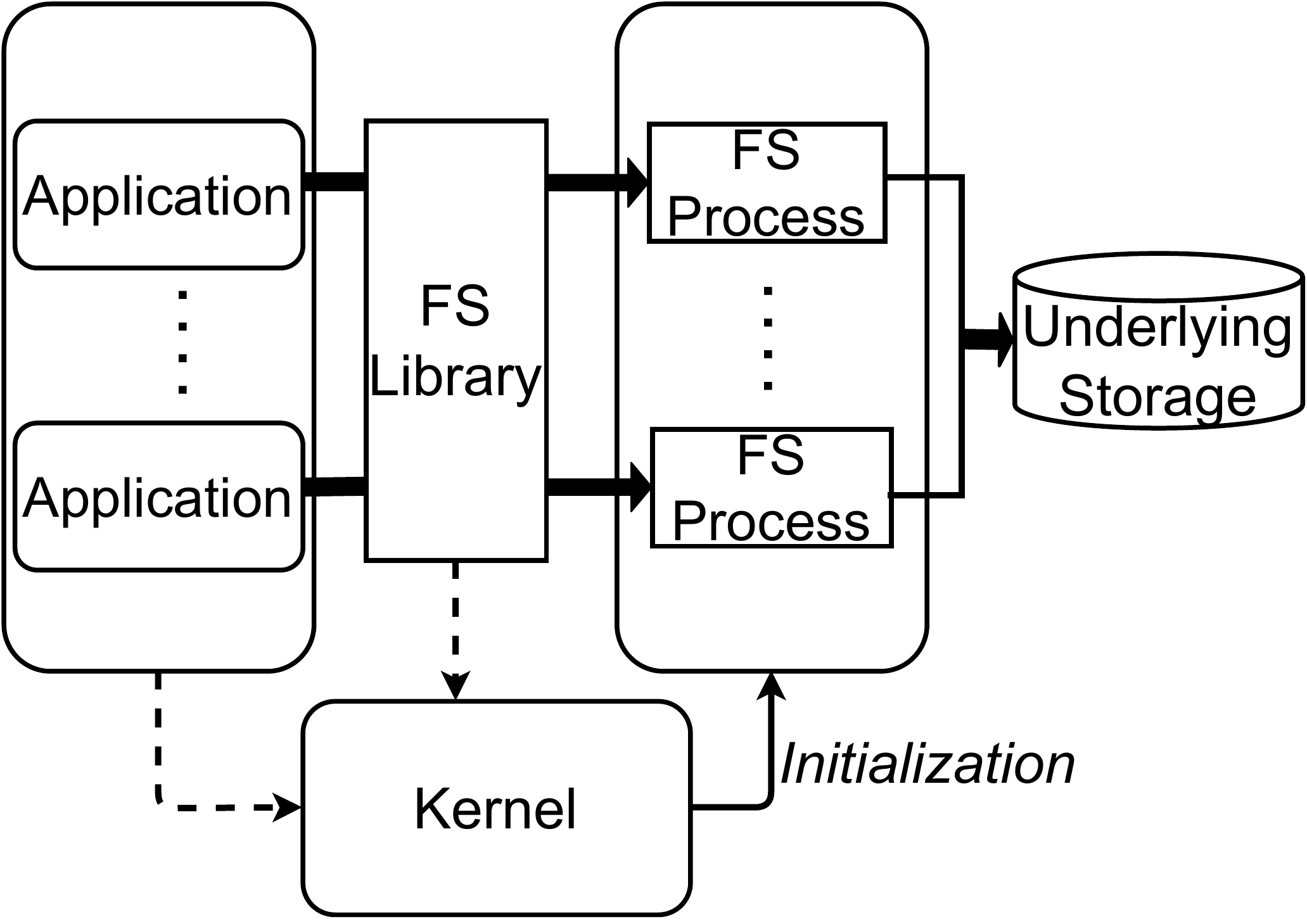}}
    \caption{Typical designs of kernel as the auxiliary plane.}
    \label{fig:auxiliary}
\end{figure*}

Instead of being constrained to the framework provided by FUSE, researchers redesigned user space file systems from scratch and adopt more aggressive modifications, so that the new file systems fit to new storage devices better. 
In this subsection, we will review multiple prior file system designs with special focus on the re-organized duties of kernel space and user space.
By examining prior file system designs with respect to the emerging storage techniques, we divide these works into two categories according to the roles of kernel. 
As shown in Figure \ref{fig:control}, several designs \cite{Aerie,Strata,splitfs,zofs} rely on kernel to maintain metadata integrity, security, consistency and etc. 
The kernel is also in charge of performing permission checks on each data request sent from the user space.
For these designs, kernel is responsible for control-plane operations.
Other designs \cite{umfs,crossfs,devfs,fsp}, as shown in Figure \ref{fig:auxiliary}, in contrast, only delegate few auxiliary responsibilities to the kernel. Namely, most frequently used functions (e.g., data read/write) are migrated to user space or device firmware while kernel serves as the auxiliary plane.

\subsubsection{Kernel As The Control Plane}
The prior work propose user-space file systems to access storage hardware directly by bypassing kernel. However, employing untrusted user library only to manage the whole file system will pose security, atomicity, and consistency issues \cite{hedayati2019hodor}. As a remedy, these file systems use kernel to handle control plane operations (e.g., metadata modifications). In this part, we chronologically review the existing file systems designs, which are under the control of the kernel.

\noindent \textbf{Sidestepping kernel for the SCM.} 
Storage-class memory (SCM), built from NVM, is a type of high-performance storage device with byte-addressability. In other words, data can be accessed directly from SCM through load/store instructions instead of traditional I/O requests. 
To utilize this characteristic, Volos et al. \cite{Aerie} presents Aerie, a file system architecture that exposes storage to user-space applications and thus reduces software overheads of deep storage stacks.   
Aerie proposes two key designs including direct access to metadata and the client-server model. Based on the two design principles, Aerie constructs an architecture 
including an untrusted library (libFS) and a trusted file-system service (TFS) in user space and an SCM manager in kernel space. 
libFS functions as the client, which has direct access to SCM when performing normal reads and writes or metadata reads. When accessing protected data or changing metadata, it needs to query the server, TFS. TFS manages protected data and metadata in user space, which is hidden from from the user applications. TFS has to query SCM manager for privileged operations. The SCM manager in kernel space then controls the privileges of SCM in storage allocation, address mapping, and extent protection.
This client-server model provides direct access to SCM while maintaining the metadata security. However, the SCM management relies on a particular hardware module (i.e., memory controller), which in turn reduces the potability to other storage devices. Besides, Aerie introduces extra latency overheads due to the query process of the client-server model.


\noindent\textbf{Designs for multi-level storage media.} Apart from stressing on one single storage technique, Kwon et al. \cite{Strata}
proposes Strata as a cross-media file system, which leverages the advantages of various storage media (i.e., NVM, SSD, and HDD).
Strata reorganizes duties of the user space and kernel space to collaborate with different storage devices, and stores different data and metadata in the most suitable media. 
At the core idea, Strata maintains per-process user-level logs in NVM to record write requests and uses kernel to \emph{digest} these updates. \emph{Digestion} is a special technique proposed in Strata. 
It is an asynchronous and periodical operation that applies transactions in the log to a kernel-managed shared area, which is built on diverse storage devices. 
With \emph{digestion}, private modifications in Strata are visible to other processes. 
Strata further improves the performance of the storage system is to leverage the locality within the multi-level storage media. Specifically, Strata uses data popularity (i.e., access frequency) to identify hot/cold data and migrates cold data to lower layers of the storage hierarchy while keeping hot data in higher layers.
This method assures low latency for popular data, thus increasing overall time efficiency. 
At the high level, Strata handles data plane operations (e.g., data writes and reading metadata from logs) in user space while managing control plane (e.g., metadata writes and digestion) within the kernel.   
Despite that Strata demonstrates desired I/O performance enhancement over prior works with respect to latency and throughput, intensive kernel involvement in user space accessing storage devices still introduces significant software overheads.

\noindent \textbf{Post-Strata designs on NVM.} 
Though the novel design of Strata \cite{Strata} casts a spark in the field of constructing file systems that utilize multi-level storage media, it does not fully utilize specific features of new storage techniques (e.g., byte-addressability of NVM).
Accordingly, Kadekodi et al. \cite{splitfs} presents SplitFS to fully utilize the byte-addressability feature of NVM.  
Similar to Strata, SplitFS splits file systems and responsibilities into user space and kernel space. Its novelty lies in the way such responsibilities are divided.
On the contrary of Strata, SplitFS authorizes user space libraries to handle regular files while moving all the metadata operations into kernel space.
Though using kernel to manage metadata control raises doubts on the efficiency of I/O requests, SplitFS brought multiple benefits. By reusing a mature and robust PM file system (ext4-DAX) in kernel for all the metadata operations, it can reduce the implementation complexity, utilize the existing features, and obtain consistency guarantees.


\noindent\textbf{Combining Aerie and Strata.}
To provide user space with direct access to storage devices, Aerie \cite{Aerie} adopts the idea of the client-server model and Strata \cite{Strata} focuses on the user-kernel collaboration.
Chen et al. \cite{kuco} combined the ideas of these two pioneer works in their hybrid file system.
The file system, named Kuco, utilizes a user-space library named Ulib as the client to provide unified POSIX interfaces, and a trusted kernel thread called Kfs as the server to handle requests from Ulib.
The user-kernel collaboration mechanism in Kuco offloads most tasks to Ulib and few to Kfs, avoiding performance bottleneck caused by being trapped into the kernel. Kfs mainly handles metadata updates, in which Ulib pre-locates addresses where KFS updates metadata.
Thus, the server (Kfs) needs not wait for the client to update metadata, reducing the latency overheads brought by the client-server model as observed in Aerie. 

\subsubsection{Kernel As The Auxiliary Plane}
Although aforementioned file systems are already able to handle basic read/write requests under user mode, they still trap into kernel for control-plane management. In other words, these designs cannot completely bypass kernel, which inspires more radical optimizations. In this part, we take a close look at such file systems which further reduce kernel's involvement, in other words, kernel is auxiliary.  

\noindent \textbf{Kernel into the backstage.} 
Considering that the kernel is still involved in most operations in current user-space file systems, Chen et al. \cite{umfs} presents a user-space file system design, named UMFS, to reduce the kernel's involvement. 
UMFS exploits a contiguous virtual memory space and hardware MMU to expose files directly to user space, and designs a new mechanism named user-space journaling to guarantee crash consistency. The duty of kernel is limited to handling physical memory mapping, mounting UMFS during initialization, and managing hardware privileged operations during I/O requests. Operations that do not require hardware privileges are still kept in user space. 
Through these approaches, UMFS is capable of hiding kernel into the backstage and achieving direct access to storage in user space when serving I/O requests. 

Unlike UMFS, which depends on hardware (i.e., MMU) to minimize kernel's involvement, Dong et al. \cite{zofs} presents a method of using software to achieve such a goal, along with a user-space file system named ZoFS.
ZoFS proposes a new management unit called \emph{coffer}, an abstraction of isolated NVM pages that can store files with the identical permission. In ZoFS, user-space libraries take full control of NVM within a coffer. The kernel only guarantees cross-coffer isolation, responsible for handling metadata of coffers and coffer-level requests from user space.

\noindent \textbf{In-storage file system.} Limiting kernel's involvement for kernel-bypass while still keeping the file system in kernel space, as in UMFS, can not fully reduce the overheads introduced by kernel. 
To address such defects, Kannan et al. \cite{devfs} proposes DevFS, which deserts aforementioned design ideas where file systems are kept in kernel space, such as kernel hiding in UMFS \cite{umfs} and trusted server in Aerie \cite{Aerie}.
DevFS takes a brave step forward by integrating device firmware into file system design and bypassing kernel for control-plane operations. 
Specifically, DevFS constructs a firmware file system that resides in the storage hardware, which utilizes device-level DRAM and CPU cores. 
With the hardware directly exposed to file system, the hardware features such as device-level capacity are visible to users.
The on-device file system also exposes standard POSIX interfaces to user-space applications. 
Therefore, user applications can directly access storage devices through DevFS without trapping into kernel.
By such design, the performance of DevFS surpasses prior works by providing a true direct-access file system, which minimizes the kernel's involvement.
However, the reliance on hardware assistance, such as device-level CPUs, limits its portability to other storage systems with different device-level features. 
And the performance limits of device-level CPUs (i.e., they are typically slower than host-machine CPUs) imposes a more strict performance limit on DevFS.

\noindent\textbf{Synergizing user, kernel and firmware file systems.}
DevFS \cite{devfs} is the pioneer in constructing firmware-level file systems, meaning that they exploit firmware-level features to build user-accessible file systems. 
Such file systems are usually referred as Firmware-FS.
Nevertheless, Firmware-FS has not fully utilized the multi-core parallelism of its host system. 
Besides, both user-level file system (User-FS) and kernel-level file system (Kernel-FS) cannot achieve fully direct access.
Observing such shortcomings of current file system designs, Ren et al. \cite{crossfs} proposes CrossFS, a synergistic design that disaggregates the file system across the user-level (LibFS), OS layer, and device firmware (FirmFS), thereby taking advantages of each layer and optimizing the overall performance. 
To achieve fine-grained concurrency, CrossFS utilizes the file descriptor (rather than the \texttt{inode}) as the basic synchronization unit and assigns each file descriptor an I/O queue (named FD-queue) for requests submission. 
By such abstraction, CrossFS dispatches different file system functions to different layers. 
The user-level LibFS, using host resources (e.g., host CPU), provides unified POSIX semantics to user applications and converts received POSIX system calls to FirmFS-dedicated I/O commands. 
Then, FirmFS uses on-device resources (e.g., device DRAM and CPUs) to fetch commands from FD-queues and apply them to the underlying hardware. 
FirmFS is also responsible for metadata management, data journaling, and permission check. 
On the other hand, the duty of the OS layer is limited to FD-queue initialization, file system mounting and garbage collection, which are rarely used.
By disaggregating file system across different layers and utilizing both host and device resources, CrossFS achieves overall performance enhancement over single User-FS, Kernel-FS, and Firmware-FS.

\noindent\textbf{File system as user process.} Opposite to DevFS \cite{devfs} and CrossFS \cite{crossfs}, Liu et al. \cite{fsp} is not content with firmware file systems for the limited on-device computing resources. 
They put forward a novel idea called file systems as processes (FSP). 
FSP constructs a file system which runs as a standalone user process and invokes kernel only during its initialization. 
Replacing the role of kernel, FSP becomes the mediator between user applications and storage devices. 
FSP provides channels consisting of a control plane and a data plane to user applications, allowing applications to issue I/O requests to storage devices via these channels. 
During the I/O processing, FSP takes charge of the request handling and ensures most kernel properties (e.g., metadata integrity and crash consistency) using these channels. 
Therefore, FSP reduces kernel overheads in delivering file system services without extra hardware assistance.

\section{Application}
\label{application}
\noindent 
In this section, we give a review of the applications that are either inspired by new storage devices or deployed in modern user-space storage systems. 

\noindent\textbf{More productive equipment for developers.}
Inspired by FUSE, Schuhknecht et al. \cite{ruma} proposes a new approach named RUMA, which aims to manage physical memory allocation in user space.
Considering that efficient and secure memory management is crucial for developing data-intensive systems, RUMA claims that traditional methods for memory allocation cannot achieve both flexibility and access performance. To address this problem, without modifying the kernel, RUMA provides a user space toolset to manipulate the virtual to physical mapping, which exposes the physical memory to user space. Therefore, programmers can allocate continuous memory in physical address space, which is also available for dynamic adjustment. Similarly, Breeze \cite{breeze} also focuses on enabling developers to write efficient codes while considering NVM in its design. The directly accessible, low-latency and byte-addressable properties of NVM offer a wide range of benefits to users. However, it is demanding for programmers to fully exploit these benefits since it requires a comprehensive understanding of NVM. Aiming to make NVM programmer-friendly, Breeze launches a toolchain that includes a user-level library and a C compiler allowing programmers to write NVM-oriented codes without having particular knowledge of NVM.

\noindent\textbf{Data-intensive storage application.} The descending of the big-data era and the explosion of deep learning have promoted the research of data-intensive computing systems with new emerging storage devices \cite{caulfield2010understanding}. Besides storing data, researchers are also concerned with how to effectively access them. Han et al. \cite{Han2017AcceleratingAB} propose a novel user-level I/O framework in high performance computing (HPC) systems that implements user-level I/O isolation by leveraging multi-streamed SSDs \cite{kang2014multi}. HyCache \cite{hycache} focuses on imbalanced performance between HDDs and SSDs in distributed file systems. By developing a middleware layer to mediate upper-level distributed file systems and lower-level storage devices, it proposes a user-level file system design which manages diverse storage devices (e.g. HDDs and SSDs) and leverages their properties for performance improvement. 
Davram \cite{davram} observes the lack of systematic and efficient memory allocation mechanism in distributed big data systems. To address this challenge, Davram proposes a user-level memory management middleware, which enables non-privileged users to access distributed virtual memory without development and performance overheads.
It manages the data swapping between persistent storage and transient memory, and exposes low-level memory design details to users, thereby achieving user-level access to low-level memory. DLFS \cite{deeplearning}, or Deep Learning File System, notices the inefficient resource arrangement of deep learning applications on HPC systems while considering the high performance of NVMe devices. It then proposes a new user-level file system that disaggregates storage across NVMe devices and provides efficient methods to handle metadata and I/O services. The evaluation of DLFS shows significant performance enhancement with much less CPU utilization compared to conventional file systems in deep learning application.


\section{Conclusions and Future Work}
\label{conclusion}

The deployment of the emerging storage devices in the existing memory hierarchy has significantly increased the I/O bandwidth and reduced the I/O latency. However, this technique shift imposes huge demands for the evolution of the traditional operating systems. This challenge inspires researchers to explore user-space storage system designs, which strive to remove the involvement of kernel from data path as much as possible and maintain the consistency and security of the entire operating system. 
In this paper, we review the former research in the past decade that target towards addressing this challenge and summarize them as a survey. Specifically, we categorize the prior work into different system layers and user-level applications. In each category, we compare different works in terms of similarities and differences. 

\ifCLASSOPTIONcaptionsoff
  \newpage
\fi



%
\bibliographystyle{IEEEtran}
\bibliography{refs}



%








\end{document}